\documentstyle[floats,twocolumn,aps,psfig,prl]{revtex}
\begin{document}
\draft
\twocolumn[\hsize\textwidth\columnwidth\hsize\csname @twocolumnfalse\endcsname
\title{$\kappa$-(BEDT-TTF)$_2$Cu[N(CN)$_2$]Br: a Fully Gapped
Strong-Coupling Superconductor}
\author{H. Elsinger$^1$, J. Wosnitza$^1$, S. Wanka$^1$, J. Hagel$^1$,
D. Schweitzer$^2$, W. Strunz$^3$}
\address{
     $^1$Physikalisches Institut, Universit\"at Karlsruhe,
     76128 Karlsruhe, Germany\\
     $^2$3. Physikalisches Institut, Universit\"at Stuttgart,
     70550 Stuttgart, Germany\\
     $^3$Anorganisch-Chemisches Institut, Universit\"at Heidelberg,
     79120 Heidelberg, Germany}
\date{\today}
\maketitle

\begin{abstract}
High-resolution specific-heat measurements of the organic
superconductor
$\kappa\mbox{-(BEDT-TTF)}_2\-\mbox{Cu[N(CN)}_2\mbox{]Br}$ in the
superconducting ($B = 0$) and normal ($B = 14$\,T) state show a
clearly resolvable anomaly at $T_c = 11.5$\,K and an electronic
contribution, $C_{es}$, which can be reasonably well described by
strong-coupling BCS theory. Most importantly, $C_{es}$ vanishes
exponentially in the superconducting state which gives evidence
for a fully gapped order parameter.
\end{abstract}


\vskip2pc]

Since the discovery of superconductivity in organic metals about
20 years ago the question on the nature of this state is one of
the most intriguing problems in this class of materials. The
close neighborhood of antiferromagnetically ordered states in
the pressure-temperature phase diagram has spurred speculations
on a Cooper-pair coupling which is mediated by antiferromagnetic
fluctuations rather than by conventional electron-phonon coupling
\cite{mck97,schma98}. This notion gained additional feedback by the
growing evidence for unconventional behavior of the high-$T_c$
cuprates and heavy-fermion superconductors. A large number of
experiments, especially on the quasi-two-dimensional (2D) organic
materials, were initiated to elucidate the question on the
symmetry of the order parameter, i.e., on the determination
of possible gap nodes in the superconducting state. The outcome
is rather controversial with an approximately equal distribution
of reports which present results in line with conventional BCS-like
behavior and others giving support for an unconventional state
\cite{ish98,wos99,sch99}. Here, the term `unconventional superconductivity'
is used to denote that either a non-phononic Cooper-pair attraction
is present or that besides the gauge symmetry additional symmetries
are broken at $T_{\rm c}$.

The most studied family of the 2D organic charge-transfer salts
is the $\kappa$-phase based on the donor molecule BEDT-TTF
(bisethylenedithio-tetrathiafulvalene or ET for short).
Materials of this phase reveal a unique phase diagram
\cite{wzi96,kan97} with $\kappa$-(BEDT-TTF)$_2$\-Cu[N(CN)$_2$]Br,
the superconductor with the highest transition temperature
($T_c = 11.5$\,K) in this class, being close to an antiferromagnetic
(presumably) Mott-insulating ground state. This direct neighborhood
of competing ground states strongly motivated the speculations
on a non-phononic pairing mechanism.

Results especially in favour for unconventional behavior were supplied
by $^{13}$C-NMR experiments of $\kappa$-(BEDT-TTF)$_2$\-Cu[N(CN)$_2$]Br
\cite{des95,may95,kan96}. The NMR data were obtained with the
necessarily applied field along the BEDT-TTF planes. For this field
orientation it is believed that the vortex lattice is trapped in the
so-called lock-in state and that one thereby can avoid additional
spin-relaxation processes due to the otherwise present
flux-line motion. All three
experiments \cite{des95,may95,kan96} showed consistently a
non-exponential, i.e., non-BCS-like, decrease of the spin-lattice
relaxation rate $1/T_1$. The data could approximately be described
by a $1/T_1 \propto T^3$ dependence which was interpreted as an
indication for $d$-wave pairing with line nodes in the energy gap.
Accordingly, these line nodes should lead to a $T^2$ behavior of
the electronic specific heat in the superconducting state, $C_{es}$.
Recently, indeed specific-heat data were reported \cite{nak97} which
seemingly showed an approximately $T^2$ dependence of $C_{es}$.
In that experiment, however, the phonon specific heat of
$\kappa$-(BEDT-TTF)$_2$Cu[N(CN)$_2$]Br was tried to estimate by
measuring a quench-cooled non-superconducting deuterated sample
which is just on the insulating side of the above-mentioned
phase diagram \cite{wzi96,kan97}.

Specific-heat experiments are an especially powerful method in
order to decide whether nodes of the superconducting gap are
present or not. If this integral technique reveals an exponential
dependence of $C_{es}$, nodes of the order parameter, i.e., points
where the superconducting gap becomes zero, can unequivocally be ruled
out. On the other side, care has to be taken when a non-exponential
behavior of $C_{es}$ is observed. Besides the existence of gap
nodes, spurious effects like a not completely superconducting
sample or an improper subtraction of non-electronic
specific-heat contributions may lead to wrong conclusions.
This experiment, i.e, the measurement of the specific heat of
one single crystal $\kappa$-(BEDT-TTF)$_2$Cu[N(CN)$_2$]Br both
in the superconducting ($B = 0$) and in the normal state at a
magnetic field of 14\,T, was initiated in order to obtain a
definitive answer to the possible existence of gap nodes
in a reliable way.

Care was taken to reduce the heat capacity of the sample holder.
This enabled us to measure one single crystal of 3.26\,mg which
contributed 50-70\% to the total heat capacity. The heat capacity
of the empty sample holder, which consists of a sapphire plate with
a thin manganin wire (20\,$\mu$m diameter) as heater and a
RuO$_2$ resistor as thermometer, was measured in all relevant fields.
The RuO$_2$ thermometer which shows in the experimental range
only a small field dependence was calibrated in fields up to 14\,T
in steps of 1\,T. The specific heat was measured in a $^4$He cryostat
equipped with a 14\,T superconducting magnet by the quasi-adiabatic
heat-pulse technique. The temperature resolution of about $\Delta T /
T < 1\cdot10^{-5}$ prevents any rounding effects at the transition
due to the experiment.

\begin{figure}[bt]
  \centerline{\psfig{file=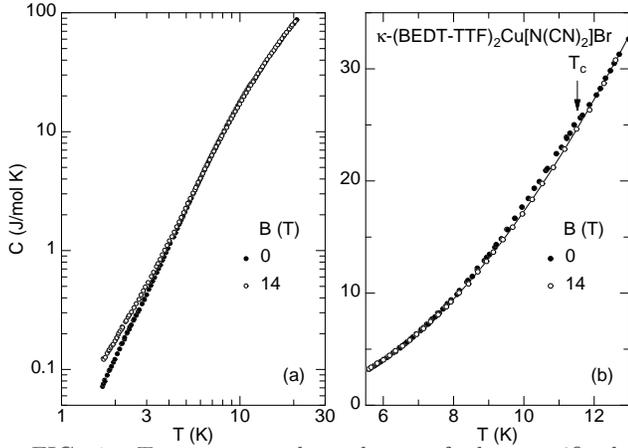,width=9cm}}
\caption[lncvslnt]{Temperature dependence of the specific heat of
$\kappa$-(BEDT-TTF)$_2$Cu[N(CN)$_2$]Br in the superconducting
($B = 0$) and normal ($B = 14$\,T) state shown (a) for the
complete temperature range and (b) for the region close to
$T_c = 11.5$\,K. The solid line in (b) is a polynomial fit
to the 14\,T data.}
\label{cvst}
\end{figure}

The specific heat, $C$, between 1.7 and 21\,K in $B = 0$ and
$B = 14$\,T is shown in Fig.\ \ref{cvst}. The upper critical field
of $\kappa$-(BEDT-TTF)$_2$Cu[N(CN)$_2$]Br is $B_{c2} = (10 \pm 2)$\,T
which can be estimated from the field dependence of our
low-temperature $C$ data (not shown) and which is in line
with earlier estimates \cite{lang,wosbook}. Therefore, the data in
$B = 14$\,T are in the normal state comprising the electronic
and the phononic contribution truely relevant for the data analysis
of this special sample. From our data we determine a Sommerfeld
coefficient $\gamma = (25\pm2)$\,mJ\,mol$^{-1}$\,K$^{-2}$ and
a Debye temperature of about $\Theta_D = (200\pm10)$\,K.
These values agree within error bars with earlier literature
data \cite{nak97,and91}. The uncertainties in our values
originate in the limited $T$ range where we observe a linear
plus a cubic temperature dependence of $C$. Already at about
3\,K we observe a deviation from the cubic Debye law, i.e.,
an additional phononic contribution. These low-lying optical
phonon modes are well known from Raman-scattering investigations
and previous specific-heat of other organic superconductors
(see Refs.\ \cite{wos94,wan98} for details).
At very low temperatures, the nuclear magnetic
moments of the hydrogen atoms of the BEDT-TTF molecules should
contribute to a Schottky anomaly due to hyperfine interactions
(see \cite{wos94} for details). In 14\,T, this hyperfine
contribution would be about 3.5\% of the total specific heat at
2\,K. In our experiment as well as in \cite{and91} no indication
of a low-temperature upturn of the $C$ data was observed for this
field. This is most probably caused by a too long spin-lattice
relaxation time compared to the thermal relaxation time of the
sample to the bath.

\begin{figure}[bt]
  \centerline{\psfig{file=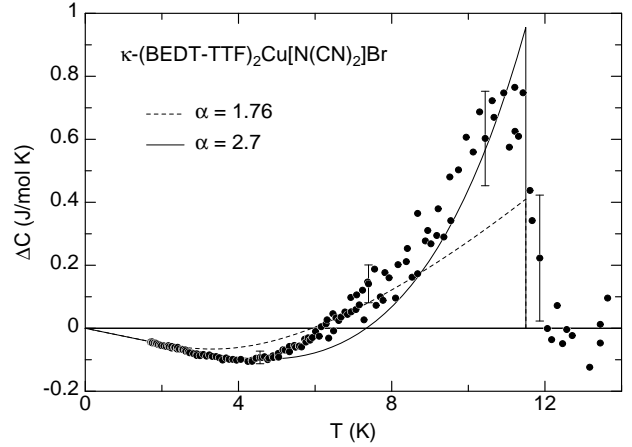,width=8.5cm}}
\caption[dcvst]{Specific-heat difference between the superconducting
and normal state with BCS curves for weak (dashed line) and strong
(solid line) coupling.}
\label{dcvst}
\end{figure}

The blow up in Fig.\ \ref{cvst}(b) shows the region close to
$T_c = 11.5$\,K. In this scale one can see more clearly the broad
anomaly arising from the superconducting transition. In contrast
to previous reports \cite{and91,kop93} we were able to unequivocally
resolve this anomaly which contributes about 3\% to the total
specific heat. The broadened jump at $T_c$ is much larger than
anticipated from weak-coupling theory. This becomes much clearer
when we plot $\Delta C$ vs $T$ (Fig.\ \ref{dcvst}),
where $\Delta C$ is the specific-heat difference between $C$
in the superconducting ($B =0$) and in the normal state
($B =14$\,T). The latter was approximated by a polynomial
[solid line in Fig.\ \ref{cvst}(b)]. $\Delta C$ expected from
weak-coupling BCS theory \cite{mue59} is shown as the dashed line in
Fig.\ \ref{dcvst}. It is obvious that the jump at $T_c$ as
well as the whole temperature dependence does not follow this
behavior. Instead, the experimental data can much better be described
by strong-coupling behavior (solid line in Fig.\ \ref{dcvst}). Thereby,
we assumed a BCS-like temperature dependence of the energy gap
$\Delta(T)$ scaled by one appropriate parameter, i.e., the
gap ratio $\alpha = \Delta(0)/ k_BT_c$, which is $\alpha_{BCS}=1.76$
in the weak-coupling limit \cite{pad73}.
With this simplistic assumption and $\alpha = 2.7$ we obtain the
reasonable description shown in Fig.\ \ref{dcvst}. The jump height is
reproduced quite well taken into account that we neglected any
fluctuations. In the intermediate temperature region the data
lie somewhat above the strong-coupling line, whereas at low
temperatures, where the data are most precise, perfect agreement
is found. We want to note that we did not fit the model to the
data but rather compared visually the BCS curves for different
$\alpha$ with the data. Therefore, as well as due to the
error bar in $\gamma$, the uncertainty in $\alpha$ is about
$\pm 0.2$.

For strong-coupling superconductors only phenomenological models exist
which connect the different superconducting parameters. By use of a
large set of data from conventional superconductors the approximate
relation between the specific-heat jump $\Delta C/\gamma T_c$
and $T_c/\omega_{ln}$ is known, where $\omega_{ln}$ is the average
phonon (or, more general, coupling) energy \cite{mar86}. Further on,
the value $T_c/\omega_{ln}$ is connected with the coupling strength
$\lambda$ of the superconducting charge carriers by the modified
McMillan equation. However, for strong coupling, i.e., $\lambda$
larger than about 1.5, the McMillan equation is not valid any more
and it is more appropriate to use an empirical relation between
$T_c/\omega_{ln}$ and $\lambda$ obtained from tunneling data and
presented in Ref.\ \cite{all75}. Under the assumption that the organic
superconductors can be described by the same strong-coupling theory
as conventional superconductors this leads to a very large $\lambda$
of about 2.5. This might be in line with a recent theoretical
treatment where enhanced strong-coupling features in
quasi-two-dimensional correlated electron systems are expected
\cite{cof98}.

\begin{figure}[bt]
  \centerline{\psfig{file=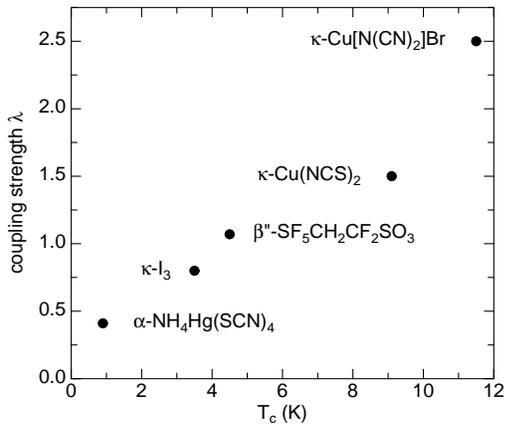,clip=,width=7cm}}
\caption[lambda]{The estimated coupling strength $\lambda$ vs
the superconducting transition temperatures for different
two-dimensional organic superconductors of the general formula
$P$-(BEDT-\-TTF)$_2X$, where the crystallographic phase $P$ and
the anion $X$ are given in the figure.}
\label{lambda}
\end{figure}

The $\lambda$ values vs $T_c$ for the
title material as well as for four other organic superconductors
\cite{wos94,wan98,gra90,nak95} are presented in Fig.\ \ref{lambda}.
Thereby, $\lambda$ was extracted for all materials in the same way,
with $\alpha = 1.76$ for the weak-coupling superconductor
$\alpha$-(BEDT-TTF)$_2$NH$_4$Hg(SCN)$_4$ \cite{nak95} and
a crudely estimated $\alpha = 2.2$ from the limited set of available
literature data for $\kappa$-(BEDT-TTF)$_2$Cu(NCS)$_2$ \cite{gra90}.
A clear systematic increase of $\lambda$, i.e., the relative
specific-heat jump $\Delta C/\gamma T_c$, as a function of $T_c$
is obvious. According to Fig.\ 1 of Ref.\ \cite{mar86} this indicates
that the characteristic average coupling energy $\omega_{ln}$
has a similar strength for all shown organic superconductors.
Consequently, one can write $\lambda \propto N(E_F)\left< I^2 \right>$
\cite{all75}, where $N(E_F)$ is the electronic density of states at
the Fermi energy and $\left< I^2 \right>$ is the coupling matrix
element averaged over the Fermi surface. Our result indicates
that mainly $\left< I^2 \right>$ controls $T_c$, since $N(E_F)$
remains more or less constant as shown by the measured $\gamma
\propto N(E_F)$ which is not correlated with Tc for the
mentioned organic superconductors. There is, however, a tendency
for a slight increase with $T_c$ if one considers only the
kappa-phase materials, from $\gamma = (18.9\pm
1.5)$\,mJ\,mol$^{-1}$\,K$^{-2}$ for $\kappa$-(BEDT-TTF)$_2$I$_3$
to $\gamma = (25\pm 2)$\,mJ\,mol$^{-1}$\,K$^{-2}$ for the title
material. Within a two-dimensional Fermi-liquid picture the
$\gamma$ values lead to effective masses of about 3.6\,$m_e$ and
4.6\,$m_e$, respectively, where $m_e$ is the free-electron mass.
This increase of $\gamma$ and the effective masses is in accordance
with results from de Haas--van Alphen or Shubnikov--de Haas
experiments which show an increasing effective cyclotron mass from
$m_c = 3.9\,m_e$ for $\kappa$-(BEDT-TTF)$_2$I$_3$ \cite{wosbook}
to $m_c = 6.6\,m_e$ for $\kappa$-(BEDT-TTF)$_2$Cu[N(CN)$_2$]Br
\cite{wei99}. These enhanced masses point to the importance of
many-body effects, i.e., electron-phonon and electron-electron
interactions, in the organic superconductors and are at least
qualitatively in line with the estimated large coupling constants
$\lambda$.

\begin{figure}[bt]
  \centerline{\psfig{file=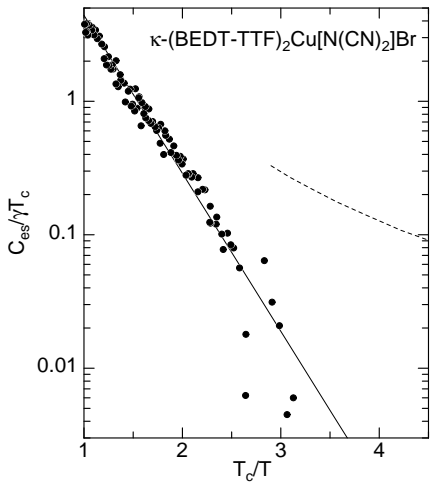,clip=,width=6cm}}
\caption[cesvsct]{Normalized plot of the electronic specific heat
in the superconducting state vs $T_c/T$. The solid line shows
the exponential vanishing of $C_{es}$. The dashed line is
the approximate estimate of $C_{es}$ from \cite{nak97}.}
\label{ces}
\end{figure}

The main point of this paper is the proof of an exponentially
vanishing electronic specific heat in the superconducting state.
It is clear already from Fig.\ \ref{dcvst} that no electronic
contribution to $C$ remains at low temperatures since otherwise
the data would not follow so perfectly the strong-coupling BCS
curve. The fact becomes more evident when we plot the
electronic part of the specific heat in the superconducting
state, $C_{es}$, as a function of $T_c/T$ (Fig.\ \ref{ces}).
For the determination of $C_{es}$ we subtracted the phonon part
of $C$ which corresponds to $C$ measured in $B = 14$\,T minus
$\gamma T$. The normalized plot in Fig.\ \ref{ces} shows
unambiguously that $C_{es}$ vanishes towards low $T$. The solid
line is an exponential fit to the data of the form $C_{es}/\gamma
T_c \propto \exp(-2.7T_c/T)$. At $T/T_c \approx 3$, $C_{es}$ is
so small that we cannot resolve it any longer leading to the scatter
of the data towards lower temperatures. From this result we can
conclude that a possible remnant of $C_{es}/T$ is less than about
1\,mJ\,mol$^{-1}$\,K$^{-2}$.
Consequently, our data prove the absence of gap nodes but, instead,
point strongly to the existence of a complete energy gap in
the superconducting state. We want to note, that our data do not
allow to make any statements on possible gap anisotropies. These
may well be the reason for the observed slight discrepancy between
the $\Delta C$ data and the BCS fit in the intermediate temperature
region shown in Fig.\ \ref{dcvst}.

Within BCS theory one can approximate $C_{es}/\gamma T_c \propto
\exp(-a_\Delta T_c/T)$ for $2.5 < T_c/T < 6$ \cite{gla69}, where
the coefficient $a_\Delta$ (= 1.44 in the weak-coupling limit) is
proportional to the energy gap $\Delta$ at $T = 0$. The much larger
value $a_\Delta \approx 2.7$ we extracted from our data is the
behavior expected for strong coupling and consistent with the
large $\lambda$. The exponential vanishing of
$C_{es}$ can equally well be proven for the organic superconductors
$\kappa$-(ET)$_2$I$_3$ \cite{wos94} and
$\beta$''-(ET)$_2$SF$_5$CH$_2$CF$_2$SO$_3$ \cite{wan98}.

In Fig.\ \ref{ces} we included the approximated average of the
estimated result for $C_{es}$ from Fig.\ 3 of Ref. \cite{nak97}
(dashed line). It is evident from our result that one can definitely
exclude any remnant contribution as high as proposed in this work
(at $T_c/T = 3$ our data are more than a factor of 10 smaller).
Indeed, the estimated $C_{es}$ at 4\,K in \cite{nak97} coincides
approximately with the normal-state electronic $C$ which would
mean a crossing of the $C$ data in the normal and superconducting
state at around this temperature. Figures \ref{cvst}(a) and
\ref{dcvst} show that this results must be wrong. It is therefore
proven that it is not allowed to estimate the phonon specific
heat from a quench-cooled non-superconducting deuterated sample.

For superconductors with line nodes a field dependence of
$\gamma$ proportional to $\sqrt{B}$ is predicted \cite{vol93}.
Recently, however, a $\sqrt{B}$ dependence was also observed at
low fields in an s-wave superconductor \cite{son99} pointing out
that the bare observation of this behavior does {\it not} prove
an unconventional pairing state. For the title material a
$\sqrt{B}$ dependence of $\gamma$ at low fields was reported
\cite{nak97}. Since our measurements were made at higher
temperatures we cannot make a definitive statement. However, from
the field dependence of $C$ at fixed temperature we can describe
the data reasonably well by a linear field dependence.

In conclusion, the results of our specific-heat measurements of
the organic superconductor $\kappa$-(BEDT-TTF)$_2$\-Cu[N(CN)$_2$]Br
in the superconducting and normal state can be well described
by strong-coupling BCS theory. We extract a large coupling
parameter $\lambda \approx 2.5$ which scales well with $\lambda$
values found for organic superconductors with lower $T_c$.
The electronic specific heat in the superconducting state
vanishes exponentially with $T_c/T$ which disproves the
$T^2$ behavior claimed earlier. Our data are fully consistent
with a completely gapped order parameter.

We thank H.\ v.\ L\"ohneysen for continuous support and fruitful
discussions. This work was partially supported by the
Deutsche Forschungs\-gemeinschaft.


\end{document}